\documentstyle[12pt]{article}
\def\ee{e^+e^-}
\def\beq{\begin{equation}}
\def\eeq{\end{equation}}
\def\beeq{\begin{eqnarray}}
\def\eeeq{\end{eqnarray}}
\def\to{\rightarrow}

\def\as{\alpha_S}

\def\gtap{\raisebox{-.5ex}{\rlap{$\,\sim\,$}} \raisebox{.5ex}{$\,>\,$}}

\def\Cpar{$C$-parameter}
\def\frac#1#2{ {{#1} \over {#2} }}
\def\rat#1#2{\mbox{\small $\frac{#1}{#2}$}}

\def\thrq{\rat 3 4}

\def\np#1#2#3{Nucl.\ Phys.\ B#1 (19#3) #2}
\def\pl#1#2#3{Phys.\ Lett.\ #1B (19#3) #2}
\def\pr#1#2#3{Phys.\ Rev.\ D #1 (19#3) #2}

\def\prl#1#2#3{Phys.\ Rev.\ Lett.\ #1 (19#3) #2}

\def\figcap{\section*{Figure Captions\markboth
        {FIGURECAPTIONS}{FIGURECAPTIONS}}\list
        {Figure 
\arabic{enumi}:\hfill}{\settowidth\labelwidth{Figure
999:}
        \leftmargin\labelwidth
        \advance\leftmargin\labelsep\usecounter{enumi}}}
 \relax

\begin{document}
\begin{titlepage}
\renewcommand{\thefootnote}{\fnsymbol{footnote}}
\begin{flushright}
     LPTHE--ORSAY 97/46 \\
     September 1997 \\
     hep-ph/9709503
\end{flushright}
\vspace*{\fill}
%\par \vskip 10mm
\begin{center}
{\Large \bf Soft-Gluon Resummation: \\
\vskip0.5ex 
a Short Review~\footnote{Research supported in part by 
EEC Programme `Human Capital and Mobility', Network 
`Physics at High Energy Colliders',
contract CHRX-CT93-0357 (DG 12 COMA).}
}\end{center}
\begin{center}
        {\bf S.\ Catani}\footnote{Invited review talk presented at
        the XXXII-nd Rencontres de Moriond, {\it QCD and High Energy
        Hadronic Interactions}, Les Arcs, France, March 1997.} \\
        INFN and Dipartimento di Fisica, Universit\`a di Firenze,\\
        Largo E. Fermi 2, I-50125 Florence, Italy\\
        LPTHE, Universit\'e Paris-Sud, B\^{a}timent 211,
        F-91405 Orsay Cedex, France
\end{center}
\par \vskip 2mm
\begin{center} {\large \bf Abstract} \end{center}
\begin{quote}

I briefly summarize some general features of soft-gluon 
contribution to inclusive cross-sections. The discussion includes the issue
of soft-gluon singularities in infrared- and collinear-safe observables.
All-order resummation and the ensuing varieties of QCD predictions are
illustrated. 

\end{quote}
\vspace*{\fill}
\end{titlepage}

\pagestyle{plain}
\renewcommand{\thefootnote}{\fnsymbol{footnote}}

\noindent {\bf 1 Introduction}
%\vskip 15pt
\bigskip

Sudakov resummation, that is, resummation of {\em double-logarithmic}
perturbative contributions produced by soft-gluon radiation, has become a topic
of broad interest in recent years. The reason is twofold.

On the theoretical side, our confidence in resummed calculations has highly
increased: resummation techniques has been extended beyond double-logarithmic
(DL) accuracy \cite{BCM} and results at leading-logarithmic (LL) and
next-to-leading logarithmic (NLL) order are available for many 
observables in different processes. In particular, using resummed
calculations to 
%next-to-leading logarithmic 
NLL accuracy one can 
consistently introduce a meaningful definition (say ${\overline {\rm MS}}$)
of the QCD coupling $\as(\mu)$, investigate their theoretical accuracy
by studying the renormalization-scale dependence and match
the predictions with full next-to-leading order (NLO) calculations thus
enlarging the applicability of perturbative QCD to wider kinematical regions
\cite{CTWT}. The list of quantities evaluated to NLL order
includes many $\ee$ shape variables 
%in the two-jet limit 
\cite{CTWT},
$Q_\perp$-distributions in the Drell-Yan process \cite{QT} and cross sections
for the production of high-mass systems via the Drell-Yan mechanism
\cite{Sterman} and in deep-inelastic lepton-hadron scattering \cite{CMW}. 
These quantities regard hard-scattering processes initiated by two
coloured partons. Results at NLL order for multi-parton procesess have 
begun to appear recently \cite{kidon}.
 
On the phenomenological side, many experimental results sensitive to 
soft-gluon resummation have become available and require detailed analyses. 
For instance, high-precision
$\ee$ data \cite{eedata}
from LEP and SLC strongly demand understanding of the Sudakov region. Last,
but not least, renewed interest on soft-gluon corrections in hadronic collisions
has been prompted by
the high-$E_T$ tail of the one-jet inclusive cross-section
measured at Tevatron \cite{jetdata}. 

Some features of soft-gluon radiation and resummation are briefly reviewed in
the following Sections.

\bigskip

\noindent {\bf 2 Soft-gluon effects in QCD cross sections}
%\vskip 15pt
\bigskip

The finite energy resolution of any particle detector
implies that physical cross
sections are always inclusive over arbitrarily-soft particles produced in the 
final state. This inclusiveness is essential in QCD calculations. Higher-order
perturbative contributions due to {\em virtual} gluons are 
infrared divergent and the divergences are exactly cancelled by radiation
of undetected {\em real} gluons. Thus, perturbative cross sections are finite 
but, precisely speaking, the cancellation does not necessarily
take place {\em order by order} in perturbation theory.
In particular kinematic configurations real and virtual contributions can be
highly unbalanced, spoiling the cancellation mechanism. As a result,
soft-gluon contribution to QCD cross sections can still be
\begin{itemize}

\item either large

\item or singular.

\end{itemize}
We comment on these points in turn.

The presence of large corrections  
that spoil the convergence of the perturbative expansion near the {\em exclusive
boundaries} of the phase space of QCD observables is well known \cite{BCM}. When
the tagged final state is forced to carry most of the total energy available
in the process (that is, one considers the quasi-elastic limit $x \to 1$, where
$x$ generically denotes inelasticity variables), the radiative tail of 
real emission is strongly suppressed, producing the loss of balance with the 
virtual contribution. Then the cancellation of the infrared divergences
%leaves a heritage in the form of
bequeaths finite higher-order contributions of the type
\beq
\label{logs}
C_{nm} \;\as^n \ln^m (1-x) \;, \;\;\; {\rm with} \;\; m \leq 2n \,,
\eeq
that can become large, $\as \ln^2(1-x) \gtap 1$, 
even if the QCD coupling is in the perturbative regime $\as \ll 1$. 
The logarithmically-enhanced terms in
Eq.~(\ref{logs}) are certainly relevant near the exclusive boundary
$x \to 1$. Moreover, their actual size in cross-section calculations
depends on the coefficients $C_{nm}$ and on the $x$-shape of parton 
densities. Thus, soft-gluon effects 
can be substantial also before reaching this extreme kinematic region.
In these cases \cite{CTWT}-\cite{kidon}, the theoretical predictions can be 
improved 
by evaluating soft-gluon contributions to high orders and possibly resumming 
them to all orders in $\as$.

%The fact that soft-gluon contribution to observables that fulfil the
%Sterman-Weinberg criteria \cite{SW} of infrared and collinear safety 
%can nonetheless be singular 
%has been pointed out recently \cite{shoulder} in general terms. 
Soft-gluon contribution can still be singular also for
observables that fulfil the
Sterman-Weinberg criteria \cite{SW} of infrared and collinear safety.
This feature has recently been pointed out \cite{shoulder} in general terms.
The singularities arise whenever the observable in question
has a non-smooth behaviour in some order of perturbation theory at an
accessible point $x_0$, called {\em critical point}, 
{\em inside the physical region} of phase space. In this case,
the lack of balance between
real and virtual contributions is just produced by the sharpness of the
distribution around the critical point $x_0$ and, 
in particular, the following `theorem' applies \cite{shoulder}.
If the distribution of the observable is 
discontinuous (e.g., it has a {\em step}) at that point in some order of
perturbation theory, 
it will become infinite
there to all (finite) higher orders. The infinite higher-order
corrections have a double-logarithmic structure similar to that in 
Eqs.~(\ref{logs}) after the replacement $\ln(1-x) \to \ln(x-x_0)$ and the
singularities can appear 
%at the critical point 
either on one side or on both sides of the critical point. 

\begin{figure}
  \centerline{
    \setlength{\unitlength}{1cm}
    \begin{picture}(0,7.5)
       \put(0,0){\includegraphics{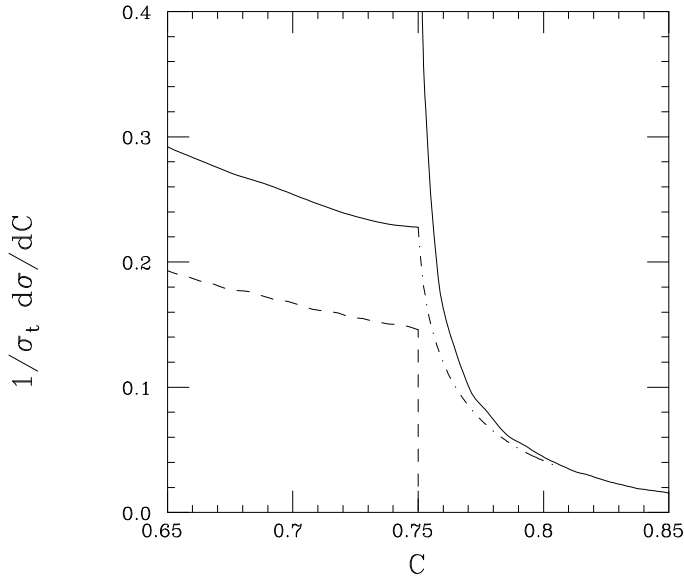}}
%       \put(0,0){\special{psfile=C3plot.ps hoffset=180 voffset=-55
%                          hscale=50 vscale=50 angle=90 }}
    \end{picture}}
  \caption[DATA]{
Predictions of the \Cpar\ distribution for $\as = 0.12$.
Dashed: ${\cal O}(\as)$. Solid: ${\cal O}(\as^2)$.
Dot-dashed: resummed. }\label{fig_Cfix}
%Contours of equal \Cpar\ for a three-particle state 
%($C_3 =$ 0.0 - 0.7).
% }\label{fig_C3}
\end{figure}

One might think that infrared- and collinear-safe
quantities, that at a certain perturbative order have a step-like behaviour 
inside the physical region, are quite abstract.
In fact, this often can happen \cite{shoulder} $i)$ if the phase-space
boundary for a certain number of partons lies inside that for a larger
number, or $ii)$ if the observable itself is defined in a non-smooth way.
The distribution of the \Cpar\ \cite{Cpar},
%,ERT}, 
a well-known event shape variable for
$\ee$ annihilation final states, is an example of type $i)$. As shown in 
Fig.~\ref{fig_Cfix}, this
distribution has a stepwise (singular) behaviour at ${\cal O}(\as)$
(${\cal O}(\as^2)$) at the value $C=C_0=\thrq$, which is the maximum value
of the \Cpar\ for three final-state partons. Observables of type $ii)$ have
typically {\em double-sided} singularities and are quite
common in jet physics. Some examples are the jet shape, 
which describes the angular distribution of energy
with respect to the jet axis \cite{CDF_D0shape,Seymour}, and the momentum
distribution of isolated photons, when the isolation criterion is defined
in terms of energy-angle cutoffs \cite{bergerphoton,frenchphoton}.

The presence of singularities in the fixed-order expansion of infrared-
and collinear-safe observables arises questions on the reliability
of perturbative predictions, on the validity (interpretation)
of the Sterman-Weinberg criteria in perturbation theory and on the effect of
non-perturbative contributions. As discussed in Ref.~\cite{shoulder} and
recalled in the following Section, these problems have a satisfactory solution
entirely within the context of perturbation theory. The solution is based
on the resummation of the singular soft-gluon contributions to all orders.

\bigskip

\noindent {\bf 3 The varieties of Sudakov resummation}
%\vskip 15pt
%\bigskip
\smallskip

\vspace{1ex}\noindent {\it Soft-gluon exponentiation}

The physical bases for all-order summation of soft-gluon contributions 
to QCD cross sections are dynamics and kinematics factorizations \cite{SCrev}.
The first factorization follows from gauge invariance 
and unitarity: in the soft limit multi-gluon amplitudes fulfil generalized 
factorization formulae given in terms of a single-gluon emission
probability that is universal (process independent). 
The second factorization regards kinematics and strongly depends on the
observable to be computed. {\em If}, in the appropriate soft limit, 
the phase-space for this observable can be written in a factorized way, 
resummation is feasible in form of {\em generalized exponentiation}
\cite{Sterman} of the single-gluon emission probability. Then, exponentiation
allows one to define and carry out an improved perturbative 
expansion that systematically resums LL terms, NLL terms and so on.

Note that phase space depends in a 
non-trivial way on multi-gluon configurations and, in general, is not
factorizable in single-particle contributions (soft-gluon exponentiation can 
be violated \cite{BS}). Moreover, even when phase-space factorization is
achievable, it does not always occur in the space where the
physical observable $x$ is defined. Usually, it is necessary to
introduce a conjugate space to overcome phase-space constraints. A typical
example \cite{Sterman,CMNT} is the energy-conservation constraint that can be 
factorized by working in $N$-moment space, $N$ being the variable conjugate to 
the energy $x$ via a Mellin (or Laplace) transformation.

Large or singular soft-gluon contributions can have
different origins (cf. Sect.~2) and resummation takes different exponentiation
forms depending on kinematics. This leads to varieties of Sudakov effects.

\vspace{1ex}\noindent {\it a) Sudakov suppression}

Soft-gluon resummation produces suppression of cross sections near the
exclusive phase-space boundary. Typical examples are $e^+e^-$ event shapes
in the two-jet limit \cite{CTWT}. Resummed predictions for these observables 
have been successfully compared with data from $e^+e^-$ annihilation 
at energies
below \cite{bZ}, at \cite{eedata} and above \cite{abZ} the $Z^0$ peak.
In particular, the predictions reduce the renormalization-scale dependence
of pure NLO calculations,
leading to measurements of $\as(M_Z)$ with smaller theoretical uncertainty.

\vspace{1ex}\noindent {\it b) Sudakov enhancement}

In hadronic collisions the exclusive phase-space boundary can be approached
in the production of systems of high mass near threshold. 
%The corresponding
Cross sections are again suppressed in this regime. However, perturbative QCD 
does not apply to absolute cross sections. The perturbatively 
computable component is what is left after factorization of 
long-distance physics into universal parton distributions. Thus, perturbative
calculations regard ratios of cross sections. In the 
%most commonly used
factorization schemes (${\overline {\rm MS}}$ and DIS)
that are most commonly used, it turns out that
soft-gluon resummation enhances the predictions for these ratios.
Moreover, parton density factorization requires exact implementation
of energy conservation. It follows that resummation must be carried out in 
$N$-moment space and the inversion to the physical space has to be performed
with care \cite{CMNT} to avoid the introduction of 
unjustified divergences \cite{sigres,BC}.

\begin{figure}
  \centerline{
    \setlength{\unitlength}{1cm}
    \begin{picture}(0,7.5)
       \put(0,0){\includegraphics{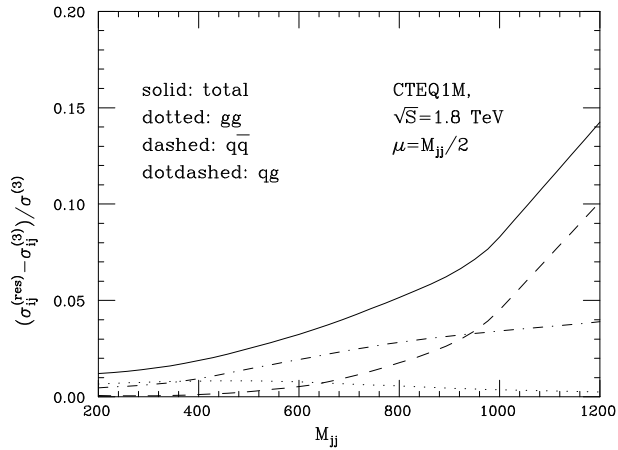}}
%       \put(0,0){\special{psfile=C3plot.ps hoffset=180 voffset=-55
%                          hscale=50 vscale=50 angle=90 }}
    \end{picture}}
\vspace*{-1.5cm}    
  \caption[DATA]{
Contribution of gluon resummation at order $\as^4$ and higher, relative to the
truncated ${\cal O}(\as^3)$ result, for the invariant-mass distribution 
of jet pairs at the Tevatron. The subscripts $ij$ refer to the various 
partonic channels.}\label{fig_jet}
\end{figure}

The effect of LL resummation on the production cross-sections of
heavy-quarks and jets was evaluated in Ref.~\cite{CMNT}. Figure~\ref{fig_jet}
shows the result for the invariant-mass $(M_{jj})$ distribution of a jet pair 
at Tevatron. Here $\sigma^{({\rm res})}$ and $\sigma^{(3)}$ respectively
denote the resummed cross-section and its truncation at order $\as^3$.
One can see that resummation enhances the cross section by 
(at most) $10\div15\%$ when the inelasticity variable $\tau=M_{jj}^2/S$ 
is as large as $\tau \sim 0.5$.

%%%%%%%%%%%%%%%%%%%%%%%%%%%%%%%%%%%%%%%%%%%
%\begin{figure}
%\vspace*{2mm}
%\centerline{\psfig{figure=jetcteqll4-p5.eps,width=0.48\textwidth,clip=}}
%\vspace*{-10mm}
%\caption{\label{jetcteq2}
%Contribution of gluon resummation at order $\as^4$ and higher, relative to the
%truncated ${\cal O}(\as^3)$ result,
%for the invariant-mass distribution of jet pairs at the Tevatron.}
%\vspace*{-5mm}
%\end{figure}
%% 

\vspace{1ex}\noindent {\it c) Sudakov shoulder}

The case of perturbative quantities with soft-gluon singularities inside
the physical phase space has, so far, received less attention in the literature.
%and a lot of work remains to be done. 
Much work on resummation remains to be done in order to reach a level of
understanding comparable to that of soft-gluon contribution  
near the exclusive phase-space boundary.

We expect \cite{shoulder} on general grounds that finiteness
will be restored at perturbative level by all-order summation.
After resummation, one obtains a characteristic structure which
is not only finite but smooth (infinitely differentiable)
at the critical point. We called this structure a 
{\em Sudakov shoulder} \cite{shoulder}.
The existence of a Sudakov shoulder implies the restoration of validity
of the Sterman-Weinberg criteria at infinite order in perturbation theory.

The dot-dashed curve in Fig.~\ref{fig_Cfix} illustrates the
Sudakov shoulder obtained by performing DL resummation
in the case of the \Cpar\ distribution \cite{shoulder}.
The resummed prediction for $C>C_0=\thrq$ joins smoothly to the $
{\cal O}(\as^2)$-calculation at (and below) the critical point.
Note that the shoulder is nevertheless still quite sharp on the 
scale shown in Fig.~\ref{fig_Cfix}.

The issue of soft-gluon singularities inside
the physical phase space is of some importance for QCD
phenomenology. Before using any ``safe'' observable to test the
theory or to measure $\as$, one needs to identify the critical
points of that observable and the expected behaviour in whatever
order of perturbation theory is to be used.  These points will
need to be avoided in comparisons between fixed-order predictions
and experiment. On the other hand, if resummed predictions can
be obtained to single-logarithmic precision, then the behaviour
at critical points represents an interesting new class of QCD predictions.

%=====

%Despite the violation of finiteness of fixed-order calculations,
%perturbative resummation suggests that non-perturbative effects
%are still power-suppressed in infrared- and collinear-safe observables.

%Much work remains to be done in order to reach a comparable level of
%understanding. 

%- soft bremsstrahlung doesn't like sharp structures
%- either: avoid them when comparing data with fixed-order perturbative
%          predictions
%- or: try to do all-order resummation to construct the Sudakov shoulder

%=======

%\noindent {\bf 4 Conclusions}
%\vskip 15pt
%\bigskip

%In summary:

%soft-gluon resummation relevant for 

%- precision tests of QCD (measurements of $\as$, ...)

%- accurate evaluation of background for new-physics signals

%- (possibly) investigations of the boundary between perturbative and
%  non-perturbative QCD (renormalons, ....)

%

%

\end{document}